# U-GIFT: Uncertainty-Guided Firewall for Toxic Speech in Few-Shot Scenario

Jiaxin Song
Beijing University of Posts and Telecommunications
Haidian Qu, Beijing Shi, China
sobgjx21@bupt.edu.cn

Xinyu Wang
Beijing University of Posts and Telecommunications
Haidian Qu, Beijing Shi, China
xywang_1@bupt.edu.cn

Yihao Wang
Beijing University of Posts and Telecommunications
Haidian Qu, Beijing Shi, China
yh-wang@bupt.edu.cn

Yifan Tang
Beijing University of Posts and Telecommunications
Haidian Qu, Beijing Shi, China
tyfcs@bupt.edu.cn

Ru Zhang
Beijing University of Posts and Telecommunications
Haidian Qu, Beijing Shi, China
zhangru@bupt.edu.cn

Jianyi Liu
Beijing University of Posts and Telecommunications
Haidian Qu, Beijing Shi, China
liujianyi1987@163.com

Gongshen Liu
Shanghai Jiao Tong University
Minhang Qu, Shanghai Shi, China
lgshen@sjtu.edu.cn

## Abstract

With the widespread use of social media, user-generated content has surged on online platforms. When such content includes hateful, abusive, offensive, or cyberbullying behavior, it is classified as toxic speech, posing a significant threat to the online ecosystem's integrity and safety. While manual content moderation is still prevalent, the overwhelming volume of content and the psychological strain on human moderators underscore the need for automated toxic speech detection. Previously proposed detection methods often rely on large annotated datasets; however, acquiring such datasets is both costly and challenging in practice. To address this issue, we propose an uncertainty-guided firewall for toxic speech in few-shot scenarios, U-GIFT, that utilizes self-training to enhance detection performance even when labeled data is limited. Specifically, U-GIFT combines active learning with Bayesian Neural Networks (BNNs) to automatically identify high-quality samples from unlabeled data, prioritizing the selection of pseudo-labels with higher confidence for training based on uncertainty estimates derived from model predictions. Extensive experiments demonstrate that U-GIFT significantly outperforms competitive baselines in few-shot detection scenarios. In the 5-shot setting, it achieves a 14.92% performance improvement over the basic model. Importantly, U-GIFT is user-friendly and adaptable to various pre-trained language models (PLMs). It also exhibits robust performance in scenarios with sample imbalance and cross-domain settings, while showcasing strong generalization across various language applications. We believe that U-GIFT provides an efficient solution for few-shot toxic speech detection, offering substantial support for automated content moderation in cyberspace, thereby acting as a firewall to promote advancements in cybersecurity.

Content Warning: This paper contains examples of harmful language.

## 1 Introduction

Social media has experienced explosive growth in recent years, with the number of global users reaching 4.33 billion by early 2021[1]. These platforms host vast amounts of user-generated content, ranging from friendly interactions to offensive and abusive posts. Some users exploit online anonymity to harass and attack individuals based on their ethnicity, gender, religion, or sexual orientation, leading to an increase in online harassment and cyberbullying [49]. Toxic comments inflict psychological harm on targeted groups [22] and can exacerbate prejudice, potentially inciting violence or criminal behavior [4]. In response to this issue, social media platforms rely on moderators to remove content that violates community guidelines [17]. However, due to the immense volume of content and the mental health impact of prolonged exposure to harmful material, manual moderation is often inadequate. Consequently, automating the detection of toxic content to assist moderators has become a crucial area of research in cyber security.

---
[1]https://datareportal.com/social-media-users

Early toxicity detection systems employ machine learning techniques and rely on rule-based approaches grounded in basic language models to automate the identification of toxic content. For example, [7] introduced the Lexical-Syntactic Feature (LSF) architecture for detecting offensive content, [20] developed a rule-based sentiment analysis model called VADER, and [43] implemented a dictionary-based system. Although these machine learning methods can extract token-level features using predefined lexical and syntactic rules, they frequently fail to account for the contextual semantic information essential for accurate hate speech detection. This limitation results in suboptimal performance and challenges in identifying implicit toxic language that lacks overtly vulgar expressions.

With the advancement of Natural Language Processing (NLP) techniques, recent studies have increasingly focused on classifying user posts as toxic or non-toxic using supervised learning models, thereby automating the detection of toxic content. These models typically consist of three key components: word embeddings, text feature extraction, and classification. The text feature extraction module plays a crucial role in identifying toxic features and detecting distributional differences within the text, while the classifier assigns labels based on the nature of the toxic statements. Research has demonstrated the effectiveness of deep learning methods in detecting toxic language on social media [24]. For instance, [54] proposed a deep neural network that integrates convolutional and gated recurrent networks to capture word sequences in short texts, achieving strong performance across various social media datasets. Meanwhile, [15] introduced a deep learning model that combines metadata with hidden patterns from tweet texts, effectively detecting various forms of abusive behavior. In recent years, Transformer-based models such as BERT [12] and RoBERTa [27], pre-trained on millions of documents through self-supervised learning, have achieved remarkable results in hate speech and offensive language detection [34]. Specifically, [31] combined BERT with multilayer perceptrons (MLP), convolutional neural networks (CNNs), and long short-term memory networks (LSTMs) for transfer learning, resulting in significant improvements in hate speech detection performance. Furthermore, [51] investigated how fine-tuning RoBERTa with a masked language modeling (MLM) step enhanced its ability to detect offensive language, yielding exceptional results.

It is important to acknowledge that while existing toxic speech detection methods can theoretically achieve satisfactory performance with extensive labeled datasets, practical applications often require thousands of instances of toxic language annotations to adequately capture the intricacies of such content. In cases where labeled data are insufficient, models may fail to fully grasp the underlying features of toxic language, leading to issues of overfitting or underfitting. Overfitting typically arises when the training dataset is too limited, causing the model to overly rely on noise and outliers, thereby exhibiting poor generalization capabilities. Underfitting occurs when the dataset is inadequate, preventing the model from effectively identifying underlying patterns and consequently resulting in suboptimal detection performance. One of the most significant challenges in the practical application of toxic speech detection is the scarcity of labeled data for training these sophisticated models [38]. Since the majority of content generated by social media users is benign and non-toxic, toxic speech occurs relatively infrequently. This leads to an imbalanced dataset where non-toxic samples significantly outnumber toxic ones, causing models to exhibit a bias toward predicting the "non-toxic" category, which substantially diminishes the effectiveness of detecting toxic speech [6]. Moreover, acquiring a substantial amount of accurately labeled toxic data is both costly and time-consuming. Annotators must meticulously review and label large volumes of content, a process that is not only labor-intensive but also psychologically taxing. Prolonged exposure to toxic text can lead to annotator burnout and potentially trigger post-traumatic stress disorder (PTSD)[2].

To address the aforementioned challenges, we propose a few-shot detection method for toxic speech that achieves substantial improvements in detection performance even when labeled data is scarce and difficult to obtain. Our approach enhances model performance by automatically screening high-quality samples from unlabeled data in network spaces, which are then used for unsupervised self-training. This process improves the effectiveness of toxic text detection. Extensive experimental results demonstrate that our proposed method outperforms existing approaches in toxicity detection within few-shot scenarios.

The main contribution of this research is:

- **Design of U-GIFT.** To address the challenge of toxic speech detection in few-shot scenarios, we introduce U-GIFT, a novel framework for uncertainty-guided firewall for toxic speech. Specifically, we developed an innovative self-training method that integrates Bayesian inference and active learning to effectively utilize the unlabeled dataset for model fine-tuning. By incorporating BNNs to assess model uncertainty, we employ variational inference and Dropout techniques to approximate the posterior distribution, thereby generating diverse prediction samples that guide the creation of pseudo labels. Furthermore, by prioritizing pseudo-labeled samples based on the uncertainty of the model prediction, we select those with high confidence and distinctive features. This approach significantly mitigates noise interference by reducing reliance on uncertain or difficult-to-classify samples, thereby enhancing detection performance under conditions of limited labeled data.
- **The proposal of the loss function.** To address the challenges associated with supervised and unsupervised self-training, we propose an enhanced loss function that incorporates sample stability weights. By quantifying the variance of pseudo-labels for each sample, we assign stability weights, thereby encouraging the model to focus on learning robust features from stable samples while reducing reliance on noisy data. This approach optimizes the cross-entropy loss function, thereby improving robustness of the model in toxic speech detection.
- **Contribution to few-shot toxic speech detection** This study demonstrates a significant enhancement in toxic speech

---
[2]https://www.bbc.com/news/technology-51245616

detection under few-shot scenarios through extensive experiments on various few-shot labeled datasets. Notably, U-GIFT achieves excellent performance even with very limited training resources, such as 5-shot samples, delivering substantial gains over the base model, with improvement of 14.92%. The proposed method shows remarkable adaptability across popular large language models, consistently surpassing baseline approaches. Additional experiments reveal that U-GIFT effectively generalizes across diverse environments, including managing imbalanced samples, operating in multilingual settings, and navigating cross-domain scenarios. Ablation studies scientifically validate the contributions of the U-GIFT components. These studies confirm that U-GIFT can serve as a robust solution for toxic speech detection tasks, capable of addressing challenges posed by diverse data distributions in cyberspace, thereby acting as a firewall to promote advancements in cybersecurity.

The remainder of this paper is organized as follows. Section 2 reviews the existing literature on toxic speech detection and few-shot learning. Section 3 provides an in-depth description of the proposed methodology. Section 4 outlines the experimental setup, presents evaluation results from multiple perspectives, and offers a thorough discussion. Finally, Section 5 summarizes the key findings and conclusions, along with reflections and outlooks on future research directions.

## 2 Related work
### 2.1 Toxic speech detection

Various forms of toxic language, including hate speech, offensive expressions, misogyny, racism, sexism, and cyberbullying, are pervasive and detrimental to the online social media environment. Extensive research has been devoted to developing automatic methods for detecting such content on social media platforms. Early studies focused on machine learning approaches to automatically identify hate speech [52], initially relying on basic rule-based methods [7][20]. Subsequently, researchers employed traditional supervised classifiers such as multinomial naive Bayes, SVM, and random forests, often in conjunction with feature extraction techniques like Bag of Words (BoW) and Term Frequency-Inverse Document Frequency (TF-IDF) [50][29]. The impact of additional features, including syntactic structures, distributed semantics (word embeddings), and user or platform metadata, was also explored [37] [5] [47]. Davidson et al. [11] introduced a large-scale dataset and utilized Logistic Regression and SVM models along with effective n-gram features for hate speech detection.

With the advancement of neural network technology, numerous deep learning-based approaches have been developed for hate speech detection. Mahajan et al. [28] introduced a comprehensive framework that integrates multiple deep learning techniques, including BiLSTM, Bi-GRU, and CNN, thereby significantly enhancing multilingual detection performance. Frameworks based on Transformer architecture [48] have demonstrated progress in detecting toxic speech [21][25]. Mou et al. [33] illustrated the efficacy of FastText and BERT in identifying hate speech by leveraging word-level semantic information and subword knowledge. The HABERTOR model [46] enhances the training efficiency and transferability of BERT-based architectures through vocabulary optimization, integration component refinement, and adversarial training. A hate speech detection dataset was introduced in [44], utilizing GPT-2 [39] for pre-training detection models. In [56], a Sentiment Knowledge Sharing (SKS) model was proposed, combining a negative word list with multi-task learning to improve hate speech detection. Multilingual hate speech detection has been achieved via transfer learning combined with pre-trained language models(PLMs) [14].

### 2.2 Few-Shot learning

It is important to highlight that nearly all the aforementioned studies rely on the robust feature extraction capabilities of neural network models. A substantial amount of data, typically numbering in the thousands, is necessary for conducting high-quality supervised training to achieve superior detection performance. Consequently, the applicability of these methods in practical scenarios is constrained. Enhancing the performance of toxicity classification tasks using a minimal amount of labeled toxic text data holds significant value, particularly in resource-scarce environments.

To enhance the effectiveness of language models in scenarios with insufficient data for specific tasks, several few-shot learning approaches leverage general task knowledge from diverse heterogeneous tasks [18]. For instance, meta-learning techniques offer an effective framework for acquiring shared learning strategies through multi-task learning. Initially, a general model initialization that captures the characteristics of numerous different tasks is established. Subsequently, the classifier for a new few-shot task is fine-tuned via several gradient descent steps based on this initial model, thereby improving adaptability to novel tasks [13]. The study by [35] assessed the feasibility of meta-learning methods for few-shot cross-lingual hate speech detection, achieving commendable results in both cross-lingual adaptability and generalization.

Existing hate speech detection models depend heavily on manual annotations. AlKhamissi et al. [1] sought to reduce this dependence by incorporating commonsense knowledge through subtask decomposition and fine-tuning on the $ATOMIC^{20}_{20}$ and StereoSet datasets, thereby improving performance in scenarios with limited labeled samples. In reference [40], the authors pretrained monolingual transformer models (MarIA and RoBERTa) using masked language modeling (MLM) on a large corpus of unlabeled tweets and then fine-tuned these models on the EXIST 2021 annotated dataset. Their approach, which included data augmentation and cross-lingual integration, enhanced detection performance for gender discrimination, particularly regarding language specificity and few-shot subtypes. Saha et al. [41] introduced RGFS, a principles-guided few-shot classification method for detecting abusive language. RGFS employs the BERT-RLT model to assign probability scores to each token in a sentence, aiding justification and incorporating manual annotation to increase the number of labeled principle samples during training.

Semi-supervised learning (SSL) enhances the performance of specific tasks by leveraging knowledge from unlabeled data. Unlike approaches that rely on extracting general task knowledge from heterogeneous tasks or require extensive manual and external resources, SSL offers a promising alternative for addressing the few-shot learning challenge. Self-training, a straightforward yet effective SSL technique, has been shown to boost model performance across various few-shot scenarios [2], making it widely applicable in numerous practical settings [26][55]. In self-training, the initial model is trained on a limited set of labeled samples, after which pseudo-labels are assigned to unlabeled data to augment the dataset for fine-tuning. This iterative process continues until the model converges. Given the abundance of unlabeled texts in few-shot toxic speech detection, this approach can significantly enhance the toxicity detection model's learning capability. Consequently, we adopt a semi-supervised self-training framework(SSF) to design our method.

## 3 Framework

Figure 1 presents an overview of the proposed U-GIFT architecture, which consists of two key components: (I) supervised training of the basic model using a few-shot labeled dataset, and (II) unsupervised self-training. The subsequent sections will provide a detailed description of each component.

**Supervised training:** The PLM, initialized with a large corpus, serves as the basic model. By fine-tuning this model with a limited set of labeled toxic and non-toxic samples, it can initially capture rudimentary features of toxic language. However, it is evident that the features derived from such a few-shot dataset are insufficient, and consequently, the basic model has not yet achieved a high level of detection accuracy.

**Pseudo-label determination:** Continue to extract and learn more salient and generalizable features of toxic text through self-training. Using the recently fine-tuned base model, we predict unlabeled samples while employing Monte Carlo (MC) Dropout to obtain uncertainty estimates from the PLM. By applying random dropout after various hidden layers in the neural network, we approximate the model output as a random sample from the posterior distribution. This approach allows us to calculate model uncertainty by determining the random mean and variance of the output samples through multiple stochastic forward passes in the network, thereby generating pseudo-labels.

**Sample Selection:** Due to insufficient feature learning in the fine-tuned basic model, the process is susceptible to significant noise and bias, leading to high uncertainty in the prediction outcomes for certain unlabeled samples. To address this issue, uncertainty estimation is employed to identify and select samples that the basic model can predict with the highest confidence, thus minimizing confusion. These selected samples are used to form a pseudo-labeled dataset that is incorporated into the self-training process. This approach enhances subsequent training phases by facilitating the extraction and learning of more robust and distinctive features, gradually reducing distributional discrepancies.

Iterate this procedure multiple times until an efficient few-shot toxic speech detection method is achieved.

### 3.1 Basic model architecture

Supervised learning methods are crucial for toxicity detection. Transformer-based PLMs have developed robust language representation capabilities by being trained on large-scale unlabeled corpora, making them suitable as basic models for extracting and learning the feature distinctions between toxic and non-toxic texts.

In real-world scenarios involving toxic speech, we often have access to a limited amount of labeled data alongside a substantial quantity of unlabeled data. We represent the labeled dataset as $D_l = \{(x_1, y_1), ..., (x_l, y_l)\}$ comprising $l$ pairs of texts $x$ and their corresponding labels $y$, where $x_i$ represents the text and $y_i$ denotes the classification label associated with $x_i$. The unlabeled dataset is denoted as $D_u$ and consists of $u$ unlabeled texts, where $l \ll u$.

For the labeled dataset $D_l$, the $i$-th token $a_i$ in each sentence $x = \{a_1, a_2, ..., a_s\}$ is initially transformed into its corresponding embedding vector $v_i$. Here, $s$ denotes the length of the sentence, and $v_i$ comprises both word embeddings and position embeddings associated with the token. Specifically, positional embeddings are appended to the word embeddings to preserve the positional information of words within the sequence. The resulting embedding vectors $V = \{v_1, v_2, ..., v_s\}$ are subsequently fed into the basic model $M_0^W$.

In the model, $H^l$ denotes the output of the $l$-th layer Transformer in the basic model, where $L$ represents the total number of layers and $1 \leq l \leq L$. The output of the language model is denoted as $H^L = \{h_1^L, h_2^L, ..., h_n^L\} \in \mathbb{R}^{k \times D_h}$, where $h_i^L$ represents the hidden state of the $i$-th token in the $L$-th layer, and $D_h$ denotes the dimensionality of the feature vector. The PLM captures contextual information from the input text through multi-layer Self-Attention mechanisms and Feed-Forward Networks, ensuring that each $h_i^L$ output encapsulates a rich representation informed by global context. Subsequently, each instance is mapped to the probability distribution over target classes via a dense layer.

Subsequently, we integrate the true label $y_i$ with the predicted label $\hat{y}_i$ for each sample and compute the loss function to optimize the model. Assuming $N_b$ represents the batch size, the loss function is formally defined as follows:

$$\mathcal{L}(\hat{y}_i, y_i) = -\frac{1}{N_b} \sum_{i=1}^{N_b} [y_i \cdot \log(\hat{y}_i) + (1 - y_i) \cdot \log(1 - \hat{y}_i)] \quad (1)$$

Based on the backpropagation and optimization of the loss function, we fine-tune and update the baisc model $M_0^W$ using a limited set of labeled samples. While the PLM initially captures low-level semantic and syntactic features, such as lexical relationships and simple contextual dependencies—that are effective for identifying basic toxic speech, these features alone are insufficient for detecting more complex and covert forms of toxic speech.

### 3.2 Semi-supervised self-training framework

We employ self-training to augment the original labeled dataset by leveraging the unlabeled data $D_u$. Specifically, we use a few-shot labeled dataset $D_l$ as the initial training set. Through this initial training, we obtain the basic model $M_0^W$ with parameters $W_0$. This basic model is then applied to the unlabeled dataset $D_u$ to generate a pseudo-labeled dataset $D_p$. Subsequently, the basic model is fine-tuned using $D_p$ to produce an updated model $M_n^W$. This iterative process continues until the convergence criteria are met.

*3.2.1 Generation of pseudo labels.* When handling unlabeled input data, the model's output is frequently susceptible to noise, leading to unreliable predictions. In such cases, a single predicted value from the model frequently fails to adequately reflect its confidence.

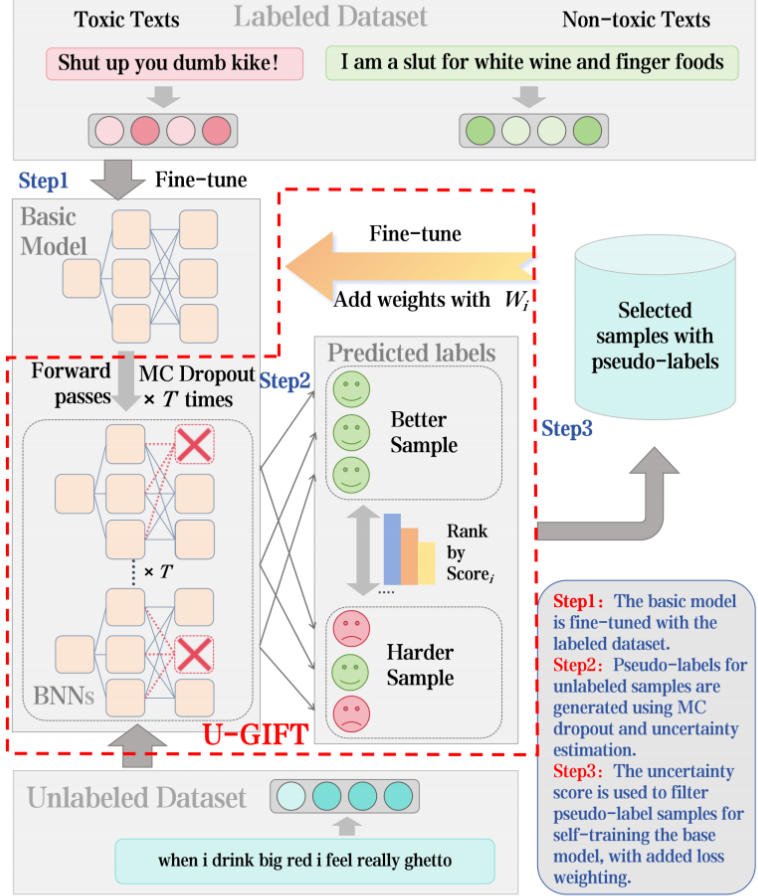

**Figure 1: Diagram of the proposed U-GIFT method, which comprises two components: the first component involves fine-tuning a basic model using labeled samples, while the second component entails self-training on unlabeled samples.**

Consequently, relying solely on deterministic outputs can result in inaccurate decisions. To address this issue, we incorporate Bayes' Theorem [16] into the basic model $M_0^W$ to provide comprehensive uncertainty quantification through probabilistic modeling of the model weights. This approach allows the model to output a probability distribution rather than a singular prediction.

In traditional neural networks, the model's output is determined by fixed weights. In contrast, BNNs assume that the model weights are not fixed parameters but rather follow a prior distribution [36]. Consequently, it is necessary to infer the posterior distribution of these weights. For a given instance $x$, we derive a probability distribution over the classes by integrating over the posterior distribution:

$$p(y = c|x) = \int_W p(y = c|M^W(x))P(W|X,Y)dW \quad (2)$$

This approach yields an evaluation of uncertainty through the aggregation of weighted averages. For classification tasks, the model output can be expressed as follows:

$$p(y = c|x, W) = \text{softmax}(M^W(x)) \quad (3)$$

Where $M^W(x) \in \mathbb{R}^h$ denotes the $h$-dimensional output generated by a Bayesian neural network, with $W$ represents the network weights, $x$ denotes the input features, and $y$ represents the class labels.

Given that the computation involved in marginalization requires summing over all possible weight configurations, it becomes prohibitively computationally intensive. Consequently, we adopt variational inference to substitute the true posterior distribution $P(W|X,Y)$ with a tractable approximate distribution $q_\theta(W)$. The optimal approximation is identified by minimizing the Kullback-Leibler (KL) divergence, thereby approximating the true posterior distribution. This approach streamlines the inference process while preserving accuracy.

We utilize Dropout as an effective technique for approximating the posterior distribution in variational inference. By randomly deactivating neurons during training, Dropout reduces model variance, enhances robustness, and mitigates the risk of overfitting. In our approach, the Dropout layer remains active not only during training but also throughout the inference phase. As a result, the model's architecture varies stochastically with each forward propagation, leading to diverse prediction outcomes. This stochasticity allows the model output to represent a distribution rather than a single deterministic value, effectively capturing the uncertainty inherent in the predictions.

To quantify this uncertainty, we perform $T$ iterations of MC Dropout, during which different neurons are randomly dropped in each iteration. We then estimate the distribution of the final output based on these varied forward propagation results. Let $q_\theta(W)$ denote the Dropout distribution, which enables us to sample $T$ masked model weights: $\{W_t\}_{t=1}^T \sim q_\theta(W)$. The approximate posterior for the classification task can be derived through Monte Carlo integration using Bayesian inference:

$$p(y=c|x) \approx \int p(y=c|M^W(x))q_\theta(W)dW \approx \frac{1}{T}\sum_{t=1}^T p(y=c|M^{W_t}(x))$$

$$= \frac{1}{T}\sum_{t=1}^T softmax(M^{W_t}(x)) \quad (4)$$

Through this approach, we can obtain multiple prediction samples for each input. These samples are used to assess the model's uncertainty and generate stable and robust predictions. Upon acquiring multiple prediction results, we derive pseudo-labels via majority voting. Specifically, for each input $x_u$ from the unlabeled dataset $D_U$, a set of pseudo-labels $\{\tilde{y}^1, \tilde{y}^2, ..., \tilde{y}^T\}$ is generated, and the most frequent prediction serves as the final pseudo-label $\tilde{y}$:

$$\tilde{y}^t = \arg\max[softmax(M^{W_n^t}(x))] = c \quad (5)$$

$$\tilde{y} = \mathbb{I}\sum_{t=1}^T \tilde{y}^t \geq \lceil\frac{T}{|c|}\rceil = c, c \in C, 1 \leq t \leq T \quad (6)$$

where $W_n^t$ represents the model parameters during the $n$-th self-training iteration at the $t$-th dropout, and $C$ denotes the predefined labels of different sample classes. $\mathbb{I}$ is the indicator function.

*3.2.2 Identification and evaluation of pseudo-Labeled samples.* The basic model we employed does not rely on external network architectures or any supplementary resources for extracting higher-level toxicity detection features. Consequently, when directly utilizing the entire unlabeled dataset or randomly selecting pseudo-labeled samples to fine-tune the basic model, significant discrepancies may arise due to errors and noise in the pseudo-labels. To mitigate this issue, it is crucial to filter the labels and select simple samples with the lowest perplexity according to the basic model and more distinct features. We perform instance sampling by incorporating model uncertainty generated through random Dropout and calculate the information gain of each sample on the model parameters, specifically focusing on the information gain between the posterior probabilities of the model:

$$E(\tilde{y}, W_n^T|x) = -\sum_c (\frac{1}{T}\sum_t \hat{p}_c^t \log(\sum_t \hat{p}_c^t)) + \frac{1}{T}\sum_{t,c}(\hat{p}_c^t \log \hat{p}_c^t), c \in C \quad (7)$$

where $\hat{p}_c^t = p(\tilde{y}_i = c|M^{W^t}(x)) = softmax(M^{W^t}(x))$. The aforementioned metric quantifies the reduction in expected posterior entropy within the output space $y$. A high value of $E(\tilde{y}, W_n^T|\cdot)$ signifies that the basic model exhibits significant uncertainty regarding the anticipated label of the instance. We employ this approach to rank all unlabeled instances based on their level of uncertainty, thereby facilitating subsequent selection for self-training purposes. The score of each sample is computed using the following formula:

$$Score_i = \frac{1 - E(\tilde{y}_i, W_n^T|x_i)}{\sum_{j=1}^U 1 - E(\tilde{y}_j, W_n^T|x_j)} \quad (8)$$

where $U$ denote the number of unlabeled samples. $Score_i$ represents the sorting weight of the $i$-th sample; a smaller value indicates that the sample is more difficult to identify. Samples with higher ranks are associated with lower model uncertainty and greater clarity regarding the expected label. By applying this filtering process, we obtain a representative subset of the pseudo-labeled dataset that is characterized by clear toxic speech features. This selected subset is then utilized for self-training to fine-tune the basic model.

*3.2.3 Training optimization incorporating stability weights.* There is an issue in unsupervised self-training where some valuable samples exhibit prediction instability, potentially causing the model to learn noise. To address this, we propose an enhanced loss function that assigns stability weights to samples based on their pseudo-label variance:

$$w_i = \alpha \cdot Var(\hat{y}_i) = \alpha \cdot \frac{1}{T}\sum_{t=1}^T \left(M^{W_n^t}(x_i) - \overline{M^{W_n^t}(x_i)}\right)^2 \quad (9)$$

Among them, $\alpha$ serves as the adjustment coefficient, $Var(\hat{y}_i)$ represents the variance of the pseudo-label for sample $x_i$. This approach encourages the model to prioritize stable and reliable samples while minimizing the impact of noisy data. Based on Equation 1, we have refined the loss function, resulting in the following optimized form:

$$\mathcal{L}_w = \sum_{i=1}^{N_b'} w_i \cdot \mathcal{L}(\hat{y}_i, y_i) \quad (10)$$

where $\mathcal{L}(\hat{y}_i, y_i)$ denotes the cross-entropy loss, and $N_b'$ s is the batch size.

## 4 Experiment and analysis
### 4.1 Settings

*4.1.1 Dataset.* To rigorously evaluate the performance of the U-GIFT method, we selected a variety of widely-used datasets from the fields of toxic speech detection, as detailed in Table 1. In both comparison and ablation experiments, benchmark datasets such as Jigsaw[8] and HateXplain[30] were utilized to conduct data segmentation, quantitative sampling. Additionally, a portion of the data underwent de-labeling operations to simulate unlabeled datasets collected from the internet. To simulate more complex multilingual environments, we also included representative non-English open-source datasets, including SWSR[23], Offenseval_2020[9], RP-Mod & RP-Crowd[3], and Korean-hate-speech[32], thereby demonstrating U-GIFT's adaptability in multilingual contexts. Furthermore, to assess U-GIFT's cross-domain detection capability, we selected datasets from various domains, including Davidson[11],

ToxiSpanSE[42], and OLID[53]. These selections ensure the robustness and effectiveness of the U-GIFT method across a wide range of application scenarios.

*4.1.2 Configuration.* PLMs like BERT, RoBERTa, and LLaMA2[45] have developed robust language representation capabilities by training on large-scale unlabeled corpora, leading to significant performance enhancements in downstream tasks. These models efficiently capture general linguistic patterns and improve both the training process and task execution efficiency through powerful pre-training initializations. Their effective regularization mechanisms help maintain strong performance while mitigating the risks of overfitting, particularly in the presence of noisy real-world data. Most of these models are based on the Transformer architecture, utilizing the attention mechanism to learn syntactic and semantic information and manage complex contextual relationships. To evaluate the detection performance of our method across different pre-trained models, we used BERT, RoBERTa, and LLaMA2-7B as classifiers in all experiments, appending by a linear layer and softmax activation. Our model implementation is conducted using PyTorch, and we applied Low-Rank Adaptation (LoRA)[19] technology from the PEFT library during fine-tuning of LLaMA2 to optimize memory usage and enhance the scalability of our experiments.

*4.1.3 Baselines.* The baselines include both classical and few-shot methods. These baselines utilize BERT and RoBERTa as backbone architectures for feature extraction in the same manner as our method during the comparison experiments. Furthermore, the training configurations are aligned with those of our method to ensure experimental consistency.

**Classic Baseline includes:** Dai et al.[10] integrated multi-task learning with BERT-based models by incorporating supervision signals from related auxiliary tasks into the BERT framework. This approach yielded performance comparable to that of the first-place winner in the English subtask A of the OffensEval competition. We consider this method as a representative example of supervised learning techniques.

**Few-shot Baselines includes:** We selected two frameworks by Saha et al.[41] for few-shot classification in detecting abusive language. These frameworks are recognized for their strong performance in toxic speech detection. The RGFS-Principles framework uses an attention mechanism to incorporate rationales (text spans justifying classification labels) into the prediction process. Specifically, RGFS-SA (RGFS-SelfAttention-Classifier) employs BERT-RLT to identify rationales, which are integrated into the BERT model via a self-attention layer. In contrast, RGFS-CA (RGFS-CrossAttention-Classifier) utilizes a cross-attention layer to incorporate rationales into the BERT model.

*4.1.4 Hyperparameters.* The framework was implemented using PyTorch on an RTX 3090 GPU. During the supervised training phase, we set the size of the labeled datasets to vary as ($K \in \{2, 5, 10, 20, 50, 100, 500, 1000\}$). Due to the small number of initial training samples, we set the batch size to 8 and performed training for 40 epochs, with the learning rate set to $3 \times 10^{-5}$. Each experiment was conducted five times with distinct random seeds and data partitions to ensure robust results. The criterion for model selection was based on the best performance over 10 iterations of self-training, with the average accuracy and F1 score reported on the blind test data. We used a test set consisting of 400 samples with a balanced 1:1 ratio of normal to toxic text.

During the self-training phase, the number of unlabeled samples was set to 10,000, with the ratio of normal to harmful text adjusted according to the experimental setup. 1,000 samples randomly selected for pseudo-labeling. We ranked these samples using the proposed method and selected the top 500 samples along with their pseudo-labels for constructing the dataset for self-training fine-tuning. To perform T=30 forward propagation steps, which is relatively time-consuming, we set the batch size to 16, conducted training for 20 epochs, and established the learning rate at $1 \times 10^{-5}$. The sample stability weight coefficient $\alpha$ was set to 0.1.

## 4.2 Overview of experiments

We provide an overview of each part of the expirements, as shown in Table 2.

## 4.3 Performance across different few-Shot sample sizes

We conducted experiments on benchmark datasets using two data sources, incrementally increasing the number of labeled examples $K$ from 2 to 1,000 while maintaining a balanced 1:1 label ratio. Analysis of the results in Table 6 indicates that as the number of labeled samples increases, both the proposed and baseline models show synchronous improvements in detection performance. This trend reflects the models' enhanced ability to extract robust statistical features from toxic text with a larger labeled dataset. A similar trend is illustrated in the radar plot in Fig. 2, where points approach the outer edge as $K$ increases.

Additionally, in various few-shot training scenarios with different values of $K$ across different toxic text datasets, U-GIFT consistently demonstrated excellent detection performance. As shown in Fig. 2, the performance representation on both sides for the different datasets reveals that the blue areas, compared to others, are more expansive, indicating that our method is effective across both benchmark datasets.

A comparison of the two plots on the left (Jigsaw) and right (HateXplain) in Fig. 2 reveals that the model performs better in toxic speech detection on the Jigsaw dataset than on HateXplain. This part aims to analyze the reasons for this observed difference. First, the Jigsaw dataset, originating from Wikipedia's discussion pages, contains comments that are generally centered around specific topics, resulting in more rational and organized content. This structured language facilitates clearer extraction of toxic language features. In contrast, the HateXplain dataset, sourced from Twitter and Gab, often features colloquial, emotional, and informal text, filled with slang, abbreviations, internet jargon, and emojis. These elements introduce irregularities that complicate hate speech detection. Moreover, comments in the Jigsaw dataset are typically longer and richer in contextual information, providing the detection model with valuable syntactic cues. In contrast, texts in the HateXplain dataset tend to be shorter and lack such context. Particularly on Twitter, character limits lead to highly condensed content, requiring the model to rely on brief clues for identifying hate speech, thereby increasing detection challenges.

Table 1: The source, total size, and language for each dataset used in the experiments.

| Datasets | Source | Total Size | Language |
| --- | --- | --- | --- |
| Jigsaw-toxic-comment[8] | Wikipedia Talk Page | 42,768 | English |
| HateXplain[30] | Twitter and Gab | 20,148 | English |
| Davidson[11] | Twitter | 24,783 | English |
| ToxiSpanSE[42] | Code Review Comments | 19,651 | English |
| Olid[53] | Twitter | 14,099 | English |
| SWSR[23] | Sina Weibo | 8,970 | Chinese |
| Offenseval_2020[9] | Twitter | 36,232 | Arabic |
| RP-Mod & RP-Crowd[3] | Rheinische Post | 57,371 | German |
| Korean-hate-speech[32] | Korean Online News | 9,381 | Korean |

Table 2: Overview of each part of the experiments.

| Type | Section | Foucs |
| --- | --- | --- |
| Intra-Dataset Performance Evaluation | 4.3 | Variations in Sample Size and Methods |
|  | 4.4 | Imbalanced samples |
| Cross-Dataset Performance Evaluation | 4.5 | Multi-language |
|  | 4.6 | Cross-domain transfer |
| Ablation expirements | 4.7.1 | SSF augmented by U-GIFT |
|  | 4.7.2 | U-GIFT |

To further validate the universality and robustness of this approach and to expand its future applicability as a plugin, we replaced the backbone network from BERT to RoBERTa and repeated the experiments using the same dataset and settings. This experiment allows for a systematic evaluation of the adaptability of different pre-trained architectures while minimizing model selection bias, with results presented in Table 7. It can be observed that the RGFS-CA and SA baseline methods experienced a decrease in performance after switching to the RoBERTa model, indicating their inability to adapt to different language models and their reliance on the original BERT. In contrast, U-GIFT showed no degradation in performance on RoBERTa; instead, it improved, achieving the best results among the baselines, thus highlighting its portability and universality across various language frameworks. We attribute this resilience of U-GIFT to RoBERTa's larger pre-training dataset, longer training duration, and greater batch size, which enhance its ability to capture features and contextual information related to toxic discourse. Additionally, advanced training methodologies and improved regularization contribute to better robustness and generalization, particularly in complex and noisy low-data scenarios.

Through a comparative analysis of the experimental results in the two tables, the U-GIFT method demonstrates superior performance over baseline methods without additional resources. In the Jigsaw dataset, U-GIFT (BERT) outperforms all baselines in a 2-label extreme few-shot scenario, achieving a significant performance gain compared to the base model. Overall, U-GIFT exhibits optimal performance with fewer than 100 samples. In the HateXplain dataset, as the dataset is expanded to 500 and 1000 samples, the Multi-Task methods are able to capture more complex semantic relationships due to the increased sample size, which slightly reduces the distinct advantages of U-GIFT. Nevertheless, U-GIFT continues to demonstrate strong performance, remaining a top contender. This highlights the limitations of existing toxic speech detection methods concerning dataset size requirements. Further analysis indicates that complex language model networks typically require substantial labeled data to learn nuanced features of toxicity effectively; without adequate support, their performance is constrained. However, our experiments reveal that many prominent toxic features can sufficiently capture distributional differences in the data, enabling effective toxicity detection even in few-shot scenarios. These findings underscore the significant advantages of our U-GIFT method in few-shot learning contexts, as it showcases remarkable performance even with extremely limited resources.

### 4.4 Performance on imbalanced datasets

Toxic content on social media platforms is often sparse, resulting in an imbalanced ratio of non-toxic to toxic texts in real-world scenarios. To more accurately simulate this environment, we adjusted the ratio of non-toxic to toxic texts in our unlabeled dataset from 2:1 to 100:1, while keeping the total volume constant. This adjustment better approximates the actual data distribution observed in practice and allows us to demonstrate the broader effectiveness and adaptability of the proposed method on the Jigsaw dataset.

To mitigate learning biases arising from limited training data and to evaluate performance across different data distributions, we opted not to use a fixed basic model for subsequent self-training. Instead, we randomly selected a small, balanced labeled sample set ($m = 10$) for initial training to fine-tune the basic model and recorded the toxicity detection results as a baseline. We then employed self-training with the imbalanced and unlabeled dataset to further enhance model performance. By comparing the changes in detection accuracy and F1 scores ($\Delta$(Acc) and $\Delta$(F1)) between the two scenarios, we assessed the performance improvements achieved through self-training when handling imbalanced unlabeled samples.

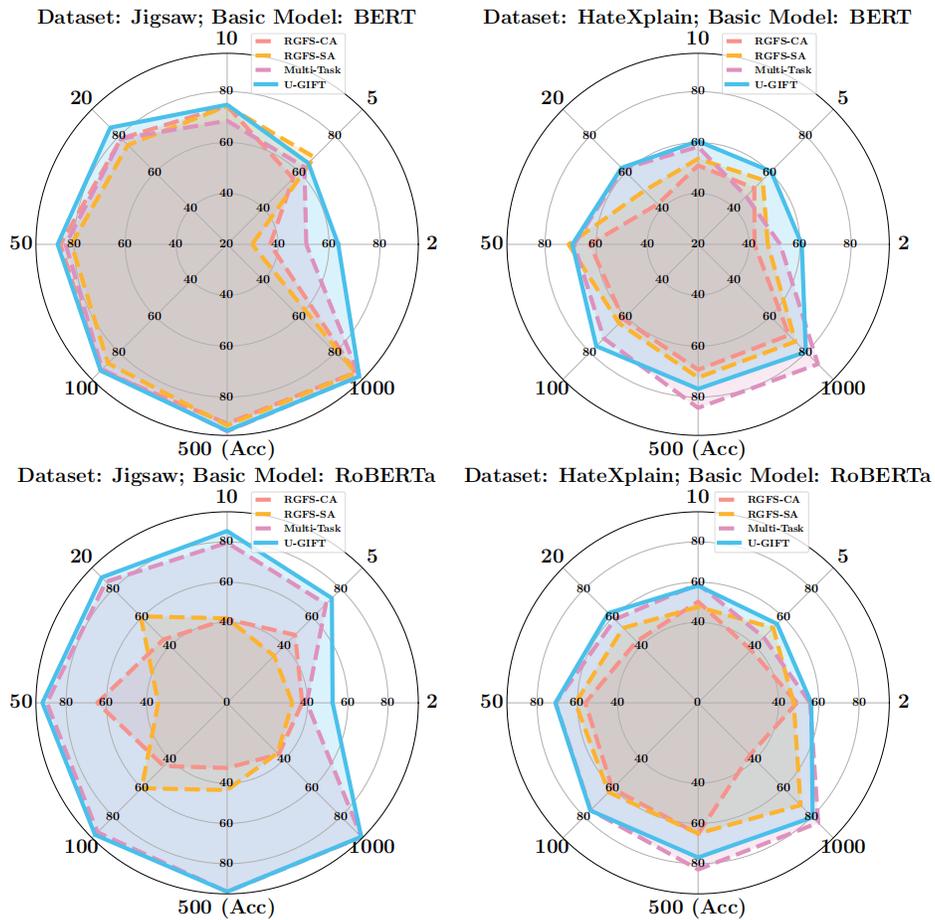

**Figure 2: Radar plots illustrating the performance of various basic models and datasets. Each subplot corresponds to a different combination of basic model and dataset. The radial axes represent accuracy (%). Points on each axis correspond to different labeled sample sizes ($K$) in the few-shot setting. A more extensive coverage area of the radar map indicates superior performance.**

Figure 3 presents the experimental results in a clear and intuitive manner. Compared to the basic model that is fine-tuned using only labeled samples, the proposed method shows substantial improvements across all experimental scenarios. The bar chart indicates a slight overall decline in detection accuracy and F1 scores as the ratio of positive to negative samples in the dataset becomes increasingly imbalanced. More notably, the line graph reveals that the performance gains of U-GIFT relative to the baseline model diminish as the imbalance in sample ratios increases. This suggests that data imbalance exerts a negative effect on model training, leading to less pronounced learned features.

As shown in Table 8, when the data ratio is set to 2:1, the accuracy and F1 score improve by 9.25% and 9.58%, respectively. This enhancement is attributable to U-GIFT's effective selection of informative samples during self-training, aided by uncertainty estimation, which mitigates the impact of unlabeled sample imbalance and bolsters the model's capacity to learn toxic features. In addition to standard imbalanced datasets, we simulated an extreme scenario where no toxic texts are present in the unlabeled dataset, resulting in a dataset comprising solely non-toxic texts (Non-toxic: toxic = 1:0). Remarkably, even under these conditions, the method remains effective, achieving a $\Delta$(Acc) of 4.25% and a $\Delta$(F1) of 4.29%. This can be attributed to the baseline model's acquisition of fundamental distinguishing features between normal and toxic texts during the initial training phase. With only normal texts available, the model can refine relevant task knowledge through self-training, thereby enhancing detection accuracy. These results not only validate the effectiveness of the proposed method in scenarios marked by a scarcity of toxic texts but also underscore its broad potential and applicability in real-world contexts.

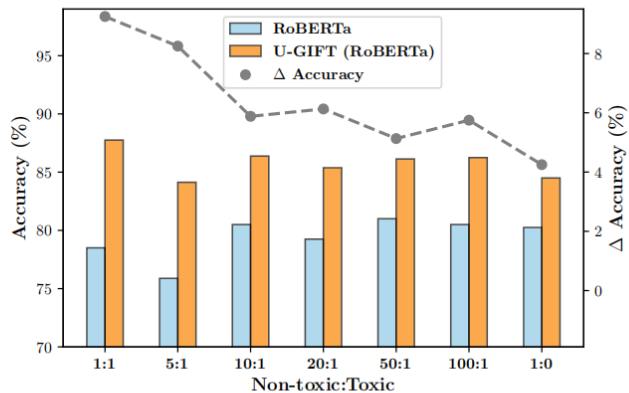

Figure 3: Performance and Variations on Different Imbalanced Unlabeled Jigsaw Datasets ($K = 10$). The bar chart represents the performance of the basic model RoBERTa (in blue) and the U-GIFT method (in orange) across varying ratios of unlabeled datasets. The line graph above illustrates the trending effect of the U-GIFT gains.

### 4.5 Performance across multiple languages

Social media applications in the real world are diverse, with user comments potentially appearing in multiple languages beyond just English, reflecting the varied linguistic backgrounds of users from different countries. To address the challenges of toxicity detection in a multilingual social media environment, we evaluated the applicability and generalizability of U-GIFT on datasets in four languages: Chinese, Arabic, Korean, and German. In the experiments, we employed the multilingual version of RoBERTa, known as XLM-RoBERTa (XLM-R), as the baseline model. We assessed the effectiveness of the U-GIFT method in few-shot settings using 50 and 20 labeled samples, respectively.

The results presented in Table 9 demonstrate significant performance improvements of the proposed method across all language datasets. Notably, U-GIFT exhibits exceptional performance on two character-based language datasets: Chinese and Korean. Chinese, as a logographic language, often conveys distinct semantic information through individual characters. XLM-R shows strong adaptability when handling such languages. In contrast, Korean is characterized by its complex morphology, featuring abundant affixes and auxiliary structures. The observed performance enhancements in these languages can be attributed to the U-GIFT method's strategy of prioritizing high-confidence and distinctly featured samples during model predictions, which effectively mitigates noise interference and maximizes the utility of unlabeled data in low-resource scenarios. Additionally, this method successfully addresses the challenges posed by the complex morphological variations and right-to-left writing system of Arabic, resulting in significant increases in both accuracy and F1 scores.

The visualizations in Fig.4 provide a clearer perspective on the results. Notably, the performance improvement on the German dataset is relatively modest, likely due to the complexity of its compound word structure, which presents significant challenges for tokenization. Furthermore, the lower representation of German in the XLM-R pre-training corpus may limit the model's ability to capture German semantics, restricting performance gains. Nevertheless, the experimental results across all datasets indicate that U-GIFT effectively captures the features of multilingual toxic language, providing a robust solution for few-shot toxicity detection in various languages.

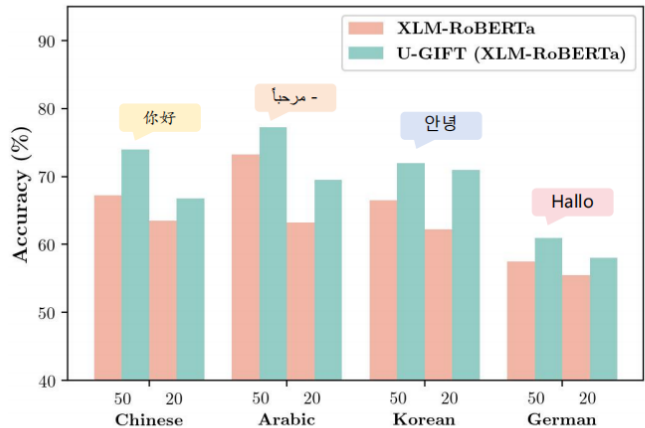

Figure 4: Performance assessment of our method on datasets in multiple non-English languages ($K = 50, 20$). The red color represents the basic model (XLM-R), while the green color indicates the U-GIFT model based on XLM-R.

### 4.6 Transferability

In practical applications of few-shot toxicity detection, the distribution of toxic texts is typically complex and diverse. Toxic texts encountered during testing may not fully align with the distribution observed in training data. This out-of-distribution (OOD) challenge places stringent demands on the robustness and adaptability of toxicity detection methods. To evaluate the transferability of our proposed method to OOD samples, we assessed the performance of a model trained on the Jigsaw dataset across three additional datasets, as detailed in the Table 3.

Table 3: Experimental results of the model trained on the Jigsaw dataset for detecting content in three OOD datasets, using $K = 20, 50,$ and $100$ as examples. The model is based on RoBERTa.

| Datasets | Metric | $K$ | | |
|---|---|---|---|---|
| | | 20 | 50 | 100 |
| Davidson | Acc | 78.75 | 72.00 | 70.75 |
| | F1 | 78.48 | 69.80 | 68.59 |
| ToxiSpanSE | Acc | 64.25 | 72.75 | 69.25 |
| | F1 | 60.78 | 72.19 | 67.24 |
| Olid | Acc | 67.25 | 66.50 | 69.50 |
| | F1 | 67.19 | 66.16 | 68.89 |

The experimental results indicate that the proposed method maintains robust detection performance on OOD datasets, as illustrated more clearly in Fig.5. On the Davidson dataset, the model achieved competitive detection performance, demonstrating its effectiveness in learning to distinguish toxic from normal text using discriminative cues from the Jigsaw dataset and successfully transferring this knowledge to datasets with different distributions. Additionally, the model exhibited strong transferability on the ToxiSpanSE and OLID datasets, which originate from diverse sources and styles. These findings suggest that the U-GIFT method displays strong adaptability in low-resource environments, effectively capturing the distributional characteristics of the data and further validating its OOD generalization capabilities.

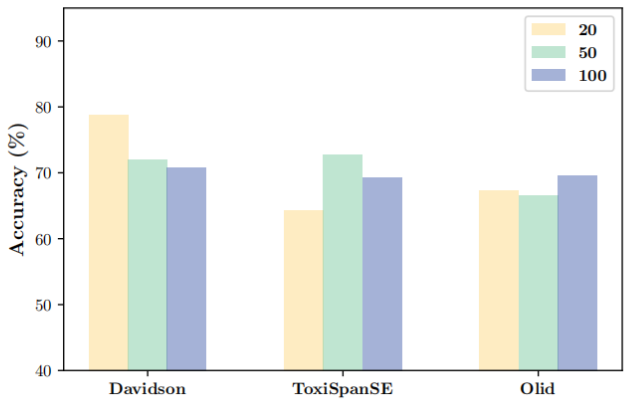

**Figure 5: Experimental results of the model trained on the Jigsaw dataset for detecting content in three OOD datasets, using $K$ = 20, 50, and 100 as examples. The model is based on RoBERTa.**

## 4.7 Ablation expirements

The elements that play a role throughout the detection process can be broadly divided into three main parts: the pre-trained language model(PLM), the semi-supervised learning framework(SSF), and the U-GIFT method integrated within them. To validate the shared methods presented in this paper, we will respectively investigate the roles of these three components.

*4.7.1 Impact of SSF(U-GIFT).* The strong performance observed in our analysis experiments can be attributed to the capabilities of PLMs. To investigate this further, we conducted ablation experiments (Table 4) using three popular models: BERT, RoBERTa, and LLaMA-2-7B. The aim of this part of the experiment is to evaluate the contributions of these models and the semi-supervised framework with U-GIFT, as two major components, to toxic speech detection.

Under extreme few-shot conditions, with only 5 labeled data points, the basic PLMs demonstrated limited performance. However, incorporating our method significantly enhanced detection accuracy across all three models. This emphasizes that the semi-supervised framework augmented by U-GIFT not only improves the performance of PLMs in few-shot scenarios but also adapts to various model architectures. Notably, RoBERTa exhibited the most significant improvement, with an accuracy increase of 14.92%, highlighting that our method effectively maximizes the potential of pre-trained language models (PLMs).

**Table 4: Ablation experiment results for detecting toxic speech under extreme few-shot conditions ($K$=5) using PLMs such as BERT, RoBERTa, and LLaMA2-7B. The table illustrates the contributions made by the SSF augmented by U-GIFT.**

|  | Metric | PLM | PLM+SSF(U-GIFT) | Δ |
|---|---|---|---|---|
| BERT | Acc | 54.83 | **60.08** | +5.25 |
|  | F1 | 48.99 | **57.09** | +8.10 |
| RoBERTa | Acc | 58.67 | **73.58** | +14.92 |
|  | F1 | 50.87 | **73.53** | +22.66 |
| Llama2-7B | Acc | 42.16 | **50.24** | +7.84 |
|  | F1 | 39.08 | **48.62** | +9.54 |

To ensure robust conclusions from our ablation experiments and avoid limiting ourselves to a specific number of samples, we conducted extensive studies across various labeled sample sizes ($K \in 5, 10, \ldots, 1000$) using LLaMA2-7B as the base model. Results in Table 10 show that the accuracy of toxic sample detection with LLaMA2-7B was notably poor with fewer than 50 labeled samples, highlighting that as a generative model, LLaMA2 struggles to adapt quickly to classification tasks under few-shot conditions. While LLaMA2 excels in language modeling, classification requires advanced feature extraction and decision boundary optimization, which typically demand larger datasets for effective tuning. Additionally, large models in few-shot fine-tuning scenarios are prone to overfitting on training data, often learning noise rather than generalizable features.

With the implementation of the proposed method, the detection performance of LLaMA2-7B shows significant improvement in training scenarios with 5, 10, 20, 50, and 100 labeled samples, as illustrated in Figure 6. Notably, with only 20 training samples, the accuracy notably exceeds that of the baseline LLaMA2-7B model. Even when the number of labeled samples increases to 1000 and LLaMA2's performance approaches its optimal level, the complete method still provides marginal but significant enhancements. These findings indicate that the semi-supervised detection method integrated with U-GIFT effectively boosts the model's performance across various few-shot learning scenarios, demonstrating robust adaptability and consistent gains. Additionally, this study preliminarily explores the integration of dropout uncertainty estimation with the LLaMA2 model, further validating the method's effectiveness and broad applicability in small sample learning environments.

*4.7.2 Impact of U-GIFT.* To thoroughly evaluate the contribution of the U-GIFT component in the semi-supervised learning framework, we designed an more specific and detailed ablation experiment aimed at comparing traditional semi-supervised self-training methods (excluding U-GIFT) with the detection method proposed in this paper that incorporates U-GIFT. To mitigate the impact of fluctuations in the supervised training results of the base model, we present its performance as a baseline in the first row of the table, indicating that the results show minimal fluctuations that can be considered negligible. The two columns of the second row show the results of the conventional self-training method without the U-GIFT component and our proposed method with the U-GIFT component, respectively. The third row provides the performance gains (delta) of both methods relative to the baseline model.

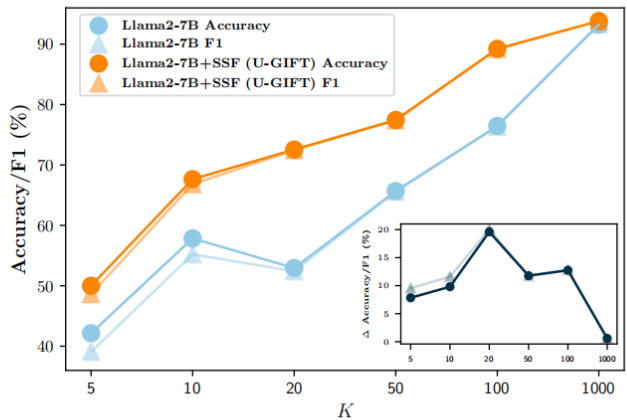

Figure 6: Performance of LLM ablation experiments for various labeled sample number ($K$), using Llama2-7B as an example, illustrates the contributions made by the SSF augmented by U-GIFT. The blue line represents the basic PLM, while the orange line shows results with the SSF augmented by U-GIFT. The black line in the lower right box illustrates the interpolation between the two approaches.

Table 5: Ablation experiment results for detecting toxic speech under few-shot conditions ($K$=20) using RoBERTa. The table illustrates the contributions made by U-GIFT.

| Framework | Metric | SSF without U-GIFT | SSF with U-GIFT | ↑ |
|---|---|---|---|---|
| PLM | Acc | 82.55 | 82.90 | \ |
| PLM | F1 | 82.44 | 82.76 | \ |
| PLM+SSF | Acc | 85.80 | **88.20** | ↑2.40 |
| PLM+SSF | F1 | 85.69 | **88.14** | ↑2.46 |
| Δ | Acc | +3.25 | +5.30 | ↑2.05 |
| Δ | F1 | +3.25 | +5.38 | ↑2.13 |

The experimental results in Table 5 indicate that the model utilizing the U-GIFT component significantly outperforms the one relying solely on conventional semi-supervised self-training for toxic text detection. Moreover, the performance improvement achieved by the U-GIFT component on the base model surpasses that of traditional self-training methods. This advantage arises primarily because conventional self-training mechanisms do not account for the uncertainty of the base model and lack effective sample selection strategies during the pseudo-labeling process. Such limitations can lead to performance drift when self-training on noisy pseudo-labeled instances. In contrast, the U-GIFT component effectively estimates the uncertainty of the base model by incorporating Bayesian networks and MC Dropout. This process not only optimizes the generation of pseudo-labels but also enhances the selection of pseudo-labeled samples by combining them with more stable sample weights, thereby improving the overall quality of the pseudo-labeled dataset. These enhancements significantly boost the model's performance in the toxic text detection task.

This ablation experiment robustly demonstrates that our U-GIFT method effectively leverages the potential of semi-supervised learning, significantly improving detection accuracy and showcasing innovation and competitiveness in the field of few-shot toxic speech detection.

## 5 Conclusion

In this paper, we propose an innovative method that integrates PLMs with self-training frameworks to address the challenge of detecting toxic content in few-shot learning scenarios. U-GIFT aims to alleviate some difficulties associated with identifying toxic speech when labeled data is scarce. Our experimental results suggest that our method can significantly outperform existing techniques even with as few as two labeled samples, indicating its potential effectiveness in extreme few-shot settings. The method is user-friendly, adaptable to various PLMs, and achieves significant performance gains across the board. Furthermore, our approach maintains high detection accuracy and F1 scores in the presence of data imbalance and multilingual tasks, highlighting its robustness and broad applicability. Additional experiments demonstrate that the U-GIFT model possesses strong generalization capabilities across various domains, further validating its performance.

Based on our findings, several promising directions for future research can be explored. First, given the potential benefits of uncertainty indicators, such as confidence scores and sample similarity, on model performance, refining and enhancing the stability weighting mechanism could facilitate more precise identification of critical samples, thereby improving model robustness. Second, incorporating multi-task learning principles and leveraging shared feature extraction approaches could further enhance model performance. Moreover, considering the increasingly multimodal nature of social media data, integrating inputs from multiple modalities—including text, images, and audio—may significantly improve model adaptability in complex environments, particularly in multimodal toxic content detection. These strategies not only extend our current research but also offer new perspectives and methodologies for tackling the challenge of toxic speech detection in practical applications.

## 6 Acknowledgments

This work is supported by the National Natural Science Foundation of China (Grant U21B2020) and supported by BUPT Excellent Ph.D. Students Foundation (Grant CX2023120).

## A  Tables of experimental results

In this section, we show the experimental results in Table 6, Table 7, Table 8, Table 9, and Table 10

## References

[1] Badr AlKhamissi, Faisal Ladhak, Srinivasan Iyer, Veselin Stoyanov, Zornitsa Kozareva, Xian Li, Pascale Fung, Lambert Mathias, Asli Celikyilmaz, and Mona Diab. 2022. ToKen: Task Decomposition and Knowledge Infusion for Few-Shot Hate Speech Detection. In *Proceedings of the 2022 Conference on Empirical Methods in Natural Language Processing*, Yoav Goldberg, Zornitsa Kozareva, and Yue Zhang (Eds.). Association for Computational Linguistics, Abu Dhabi, United Arab Emirates, 2109–2120.

[2] Massih-Reza Amini, Vasilii Feofanov, Loic Pauletto, Lies Hadjadj, Emilie Devijver, and Yury Maximov. 2025. Self-training: A survey. *Neurocomputing* 616 (2025), 128904.

[3] Dennis Assenmacher, Marco Niemann, Kilian Müller, Moritz Seiler, Dennis M Riehle, and Heike Trautmann. 2021. RP-Mod & RP-Crowd: Moderator-and Crowd-Annotated German News Comment Datasets. In *Thirty-fifth conference on neural information processing systems datasets and benchmarks track (Round 2)*.

[4] Erik Bleich. 2011. The Rise of Hate Speech and Hate Crime Laws in Liberal Democracies. *Journal of Ethnic and Migration Studies* 37, 6 (2011), 917–934.

[5] Despoina Chatzakou, Nicolas Kourtellis, Jeremy Blackburn, Emiliano De Cristofaro, Gianluca Stringhini, and Athena Vakali. 2017. Measuring# GamerGate: A tale of hate, sexism, and bullying. In *Proceedings of the 26th international conference on world wide web companion*. 1285–1290.

[6] Li-Ming Chen, Bao-Xin Xiu, and Zhao-Yun Ding. 2022. Multiple weak supervision for short text classification. *Applied Intelligence* 52, 8 (2022), 9101–9116.

[7] Ying Chen, Yilu Zhou, and Sencun Zhu. 2012. Detecting Offensive Language in Social Media to Protect Adolescent Online Safety. In *2012 International Conference on Privacy, Security, Risk and Trust and 2012 International Confernece on Social Computing*. IEEE, Heidelberg, 71–80.

[8] cjadams, Jeffrey Sorensen, Julia Elliott, Lucas Dixon, Mark McDonald, nithum, and Will Cukierski. 2017. Toxic Comment Classification Challenge. https://kaggle.com/competitions/jigsaw-toxic-comment-classification-challenge. Kaggle.

[9] Çağrı Çöltekin. 2020. A corpus of Turkish offensive language on social media. In *Proceedings of the Twelfth language resources and evaluation conference*. 6174–6184.

[10] Wenliang Dai, Tiezheng Yu, Zihan Liu, and Pascale Fung. 2020. Kungfupanda at SemEval-2020 Task 12: BERT-Based Multi-TaskLearning for Offensive Language Detection, Aurelie Herbelot, Xiaodan Zhu, Alexis Palmer, Nathan Schneider, Jonathan May, and Ekaterina Shutova (Eds.). International Committee for Computational Linguistics, Barcelona (online), 2060–2066.

[11] Thomas Davidson, Dana Warmsley, Michael Macy, and Ingmar Weber. 2017. Automated hate speech detection and the problem of offensive language. In *Proceedings of the international AAAI conference on web and social media*, Vol. 11. 512–515.

[12] Jacob Devlin, Ming-Wei Chang, Kenton Lee, and Kristina Toutanova. 2019. BERT: Pre-training of Deep Bidirectional Transformers for Language Understanding. In *Proceedings of the 2019 Conference of the North American Chapter of the Association for Computational Linguistics: Human Language Technologies, Volume 1 (Long and Short Papers)*, Jill Burstein, Christy Doran, and Thamar Solorio (Eds.). Association for Computational Linguistics, Minneapolis, Minnesota, 4171–4186.

[13] Chelsea Finn, Pieter Abbeel, and Sergey Levine. 2017. Model-agnostic meta-learning for fast adaptation of deep networks. In *International conference on machine learning*. PMLR, 1126–1135.

[14] Anderson Almeida Firmino, Cláudio de Souza Baptista, and Anselmo Cardoso de Paiva. 2024. Improving hate speech detection using cross-lingual learning. *Expert Systems with Applications* 235 (2024), 121115.

[15] Antigoni Maria Founta, Despoina Chatzakou, Nicolas Kourtellis, Jeremy Blackburn, Athena Vakali, and Ilias Leontiadis. 2019. A unified deep learning architecture for abuse detection. In *Proceedings of the 10th ACM conference on web science*. 105–114.

[16] Jakob Gawlikowski, Cedrique Rovile Njieutcheu Tassi, Mohsin Ali, Jongseok Lee, Matthias Humt, Jianxiang Feng, Anna Kruspe, Rudolph Triebel, Peter Jung, Ribana Roscher, et al. 2023. A survey of uncertainty in deep neural networks. *Artificial Intelligence Review* 56, Suppl 1 (2023), 1513–1589.

[17] Tarleton Gillespie. 2018. *Custodians of the Internet: Platforms, Content Moderation, and the Hidden Decisions That Shape Social Media*. Yale University Press, New Haven, CT.

[18] Kai He, Nan Pu, Mingrui Lao, Erwin M Bakker, and Michael S Lew. 2023. Dual selective knowledge transfer for few-shot classification. *Applied Intelligence* 53, 22 (2023), 27779–27789.

[19] Edward J Hu, Yelong Shen, Phillip Wallis, Zeyuan Allen-Zhu, Yuanzhi Li, Shean Wang, Lu Wang, and Weizhu Chen. 2021. Lora: Low-rank adaptation of large language models. *arXiv preprint arXiv:2106.09685* (2021).

[20] C. Hutto and Eric Gilbert. 2014. VADER: A Parsimonious Rule-Based Model for Sentiment Analysis of Social Media Text. *Proceedings of the International AAAI Conference on Web and Social Media* 8, 1 (2014), 216–225.

[21] Òscar Garibo i Orts. 2019. Multilingual detection of hate speech against immigrants and women in Twitter at SemEval-2019 task 5: Frequency analysis interpolation for hate in speech detection. In *Proceedings of the 13th International Workshop on Semantic Evaluation*. 460–463.

[22] Terje Olsen Janikke Solstad Vedeler and John Eriksen. 2019. Hate speech harms: a social justice discussion of disabled Norwegians' experiences. *Disability & Society* 34, 3 (2019), 368–383.

[23] Aiqi Jiang, Xiaohan Yang, Yang Liu, and Arkaitz Zubiaga. 2022. SWSR: A Chinese dataset and lexicon for online sexism detection. *Online Social Networks and Media* 27 (2022), 100182.

[24] Habibe Karayiğit, Çiğdem İnan Acı, and Ali Akdağlı. 2021. Detecting abusive Instagram comments in Turkish using convolutional Neural network and machine learning methods. *Expert Systems with Applications* 174 (2021), 114802.

[25] Brendan Kennedy, Xisen Jin, Aida Mostafazadeh Davani, Morteza Dehghani, and Xiang Ren. 2020. Contextualizing Hate Speech Classifiers with Post-hoc Explanation. In *Proceedings of the 58th Annual Meeting of the Association for Computational Linguistics*, Dan Jurafsky, Joyce Chai, Natalie Schluter, and Joel Tetreault (Eds.). Association for Computational Linguistics, Online, 5435–5442.

[26] Di Li, Xiaoguang Zhu, and Liang Song. 2023. Mutual match for semi-supervised online evolutive learning. *Applied Intelligence* 53, 3 (2023), 3336–3350.

[27] Yinhan Liu. 2019. Roberta: A robustly optimized bert pretraining approach. *arXiv preprint arXiv:1907.11692* 364 (2019).

[28] Esshaan Mahajan, Hemaank Mahajan, and Sanjay Kumar. 2024. EnsMulHateCyb: Multilingual hate speech and cyberbully detection in online social media. *Expert Systems with Applications* 236 (2024), 121228.

[29] Shervin Malmasi and Marcos Zampieri. 2018. Challenges in discriminating profanity from hate speech. *Journal of Experimental & Theoretical Artificial Intelligence* 30, 2 (2018), 187–202.

[30] Binny Mathew, Punyajoy Saha, Seid Muhie Yimam, Chris Biemann, Pawan Goyal, and Animesh Mukherjee. 2021. Hatexplain: A benchmark dataset for explainable hate speech detection. In *Proceedings of the AAAI conference on artificial intelligence*, Vol. 35. 14867–14875.

[31] Khouloud Mnassri, Praboda Rajapaksha, Reza Farahbakhsh, and Noel Crespi. 2022. BERT-based ensemble approaches for hate speech detection. In *GLOBECOM 2022-2022 IEEE Global Communications Conference*. IEEE, 4649–4654.

[32] Jihyung Moon, Won Ik Cho, and Junbum Lee. 2020. BEEP! Korean Corpus of Online News Comments for Toxic Speech Detection. In *Proceedings of the Eighth International Workshop on Natural Language Processing for Social Media*, Lun-Wei Ku and Cheng-Te Li (Eds.). Association for Computational Linguistics, Online, 25–31.

[33] Guanyi Mou, Pengyi Ye, and Kyumin Lee. 2020. Swe2: Subword enriched and significant word emphasized framework for hate speech detection. In *Proceedings of the 29th ACM International Conference on Information & Knowledge Management*. 1145–1154.

[34] Marzieh Mozafari, Reza Farahbakhsh, and Noël Crespi. 2020. A BERT-Based Transfer Learning Approach for Hate Speech Detection in Online Social Media. In *Complex Networks and Their Applications VIII*, Hocine Cherifi, Sabrina Gaito, José Fernendo Mendes, Esteban Moro, and Luis Mateus Rocha (Eds.). Springer International Publishing, Cham, 928–940.

[35] Marzieh Mozafari, Reza Farahbakhsh, and Noel Crespi. 2022. Cross-lingual few-shot hate speech and offensive language detection using meta learning. *IEEE Access* 10 (2022), 14880–14896.

[36] Subhabrata Mukherjee and Ahmed Awadallah. 2020. Uncertainty-aware self-training for few-shot text classification. *Advances in Neural Information Processing Systems* 33 (2020), 21199–21212.

[37] Chikashi Nobata, Joel Tetreault, Achint Thomas, Yashar Mehdad, and Yi Chang. 2016. Abusive language detection in online user content. In *Proceedings of the 25th international conference on world wide web*. 145–153.

[38] Fabio Poletto, Valerio Basile, Manuela Sanguinetti, Cristina Bosco, and Viviana Patti. 2021. Resources and benchmark corpora for hate speech detection: a systematic review. *Language Resources and Evaluation* 55 (2021), 477–523.

[39] Alec Radford, Jeffrey Wu, Rewon Child, David Luan, Dario Amodei, Ilya Sutskever, et al. 2019. Language models are unsupervised multitask learners. *OpenAI blog* 1, 8 (2019), 9.

[40] Francisco Rodríguez-Sánchez, Jorge Carrillo-de Albornoz, and Laura Plaza. 2024. Detecting sexism in social media: an empirical analysis of linguistic patterns and strategies. *Applied Intelligence* 54, 21 (2024), 10995–11019.

[41] Punyajoy Saha, Divyanshu Sheth, Kushal Kedia, Binny Mathew, and Animesh Mukherjee. 2023. Rationale-Guided Few-Shot Classification to Detect Abusive Language. In *ECAI 2023*. IOS Press, 2041–2048.

[42] Jaydeb Sarker, Sayma Sultana, Steven R Wilson, and Amiangshu Bosu. 2023. ToxiSpanSE: An explainable toxicity detection in code review comments. In *2023 ACM/IEEE International Symposium on Empirical Software Engineering and Measurement (ESEM)*. IEEE, 1–12.

Table 6: Experimental results of various toxic text detection methods across different numbers of labeled texts($K$) in multiple datasets, using BERT as the basic model. The Average value represents the mean of the three baseline models. The $\Delta$ value indicates the difference between U-GIFT and the Average of the baseline models. (Note: Bold entries highlight the best performance in the relevant cases discussed in the manuscript, while underlined and *marked entries indicate the second-best results. And the following is the same.)

| Datasets | Method | Metric | \multicolumn{8}{c}{$K$} |
| --- | --- | --- | --- | --- | --- | --- | --- | --- | --- | --- |
| | | | 2 | 5 | 10 | 20 | 50 | 100 | 500 | 1000 |
| Jigsaw | RGFS-CA | Acc | 36.87 | 56.60 | 74.33* | 78.80* | 85.43* | 90.10* | 90.37 | 91.13 |
| | | F1 | 48.57 | 60.67 | **74.80** | 78.90* | 85.53* | 90.13* | 90.40 | 91.13 |
| | RGFS-SA | Acc | 29.77 | **68.37** | 74.30 | 75.13 | 81.07 | 86.07 | 91.07 | 91.37 |
| | | F1 | 37.87 | **68.83** | 74.37 | 75.40 | 81.07 | 86.10 | 91.07 | 91.37 |
| | Multi-Task | Acc | 51.05* | 62.87 | 68.52 | 78.70 | 83.38 | 89.32 | **93.39** | 93.22* |
| | | F1 | 52.00* | 66.33* | 69.17 | 78.75 | 83.42 | 89.42 | **93.42** | 93.25* |
| | Average | Acc | 39.23 | 62.61 | 72.38 | 77.54 | 83.29 | 88.50 | 91.61 | 91.91 |
| | | F1 | 46.15 | 65.28 | 72.78 | 77.68 | 83.34 | 88.55 | 91.63 | 91.92 |
| | U-GIFT | Acc | **63.58** | 65.08* | **74.83** | **84.83** | **86.50** | **90.17** | 93.17* | **93.33** |
| | | F1 | **62.17** | 64.71 | 74.63* | **84.79** | **86.49** | **90.13** | 92.90* | **93.30** |
| | $\Delta$ | Acc | +24.35 | +2.47 | +2.45 | +7.29 | +3.21 | +1.67 | +1.56 | +1.43 |
| | | F1 | +16.02 | -0.57 | +1.85 | +7.11 | +3.15 | +1.58 | +1.27 | +1.39 |
| HateXplain | RGFS-CA | Acc | 42.13 | 51.20 | 50.97 | 42.37 | 62.50 | 61.50 | 69.23 | 70.40 |
| | | F1 | 44.00 | 59.57 | 51.23 | 57.20 | 63.17 | 63.00 | 71.60 | 71.00 |
| | RGFS-SA | Acc | 47.33 | 55.83* | 53.63 | 49.80 | **70.80** | 63.83 | 72.33 | 73.97 |
| | | F1 | 49.40 | 62.23 | 56.67 | 54.17 | **71.17** | 65.47 | 72.93 | 74.17 |
| | Multi-Task | Acc | 52.20* | 46.63 | 58.34* | 61.84* | 68.68 | 72.27* | **84.22** | **86.48** |
| | | F1 | 55.33* | 50.00 | 59.25* | 62.58 | 68.75 | 72.67* | **84.25** | **86.50** |
| | Average | Acc | 47.22 | 51.22 | 54.31 | 51.34 | 67.33 | 65.87 | 75.26 | 76.95 |
| | | F1 | 49.58 | 57.27 | 55.72 | 57.98 | 67.70 | 67.05 | 76.26 | 77.22 |
| | U-GIFT | Acc | **60.67** | **60.50** | **60.50** | **62.50** | 69.33* | **76.50** | 76.75* | 79.75* |
| | | F1 | **58.27** | 59.97* | **60.30** | 61.99* | 68.96* | **76.17** | 76.62* | 79.70* |
| | $\Delta$ | Acc | +13.45 | +9.28 | +6.19 | +11.16 | +2.01 | +10.63 | +1.49 | +2.80 |
| | | F1 | +8.70 | +2.71 | +4.58 | +4.01 | +1.27 | +9.13 | +0.36 | +2.48 |

[43] S. E. Smith. 2018. ElSherief, Mai and Kulkarni, Vivek and Nguyen, Dana and Wang, William Yang and Belding, Elizabeth. In *Proceedings of the international AAAI conference on web and social media*, Vol. 12.
[44] Serra Sinem Tekiroğlu, Yi-Ling Chung, and Marco Guerini. 2020. Generating Counter Narratives against Online Hate Speech: Data and Strategies. In *Proceedings of the 58th Annual Meeting of the Association for Computational Linguistics*, Dan Jurafsky, Joyce Chai, Natalie Schluter, and Joel Tetreault (Eds.). Association for Computational Linguistics, Online, 1177–1190.
[45] Hugo Touvron, Louis Martin, Kevin Stone, Peter Albert, Amjad Almahairi, Yasmine Babaei, Nikolay Bashlykov, Soumya Batra, Prajjwal Bhargava, Shruti Bhosale, et al. 2023. Llama 2: Open foundation and fine-tuned chat models. *arXiv preprint arXiv:2307.09288* (2023).
[46] Thanh Tran, Yifan Hu, Changwei Hu, Kevin Yen, Fei Tan, Kyumin Lee, and Se Rim Park. 2020. HABERTOR: An Efficient and Effective Deep Hatespeech Detector. In *Proceedings of the 2020 Conference on Empirical Methods in Natural Language Processing (EMNLP)*, Bonnie Webber, Trevor Cohn, Yulan He, and Yang Liu (Eds.). Association for Computational Linguistics, Online, 7486–7502.
[47] Elise Fehn Unsvåg and Björn Gambäck. 2018. The effects of user features on Twitter hate speech detection. In *Proceedings of the 2nd workshop on abusive language online (ALW2)*. 75–85.
[48] A Vaswani. 2017. Attention is all you need. *Advances in Neural Information Processing Systems* (2017).
[49] Emily A Vogels. 2021. *The state of online harassment*. Pew Research Center, Washington, DC.
[50] Zeerak Waseem and Dirk Hovy. 2016. Hateful Symbols or Hateful People? Predictive Features for Hate Speech Detection on Twitter. In *Proceedings of the NAACL Student Research Workshop*, Jacob Andreas, Eunsol Choi, and Angeliki Lazaridou (Eds.). Association for Computational Linguistics, San Diego, California, 88–93.
[51] Gregor Wiedemann, Seid Muhie Yimam, and Chris Biemann. 2020. UHH-LT at SemEval-2020 Task 12: Fine-Tuning of Pre-Trained Transformer Networks for Offensive Language Detection. In *Proceedings of the Fourteenth Workshop on Semantic Evaluation*, Aurelie Herbelot, Xiaodan Zhu, Alexis Palmer, Nathan Schneider, Jonathan May, and Ekaterina Shutova (Eds.). International Committee for Computational Linguistics, Barcelona (online), 1638–1644.
[52] Michael Wiegand, Josef Ruppenhofer, Anna Schmidt, and Clayton Greenberg. 2018. Inducing a Lexicon of Abusive Words – a Feature-Based Approach. In *Proceedings of the 2018 Conference of the North American Chapter of the Association for Computational Linguistics: Human Language Technologies, Volume 1 (Long Papers)*, Marilyn Walker, Heng Ji, and Amanda Stent (Eds.). Association for Computational Linguistics, New Orleans, Louisiana, 1046–1056.
[53] Marcos Zampieri, Shervin Malmasi, Preslav Nakov, Sara Rosenthal, Noura Farra, and Ritesh Kumar. 2019. Predicting the Type and Target of Offensive Posts in Social Media. In *Proceedings of the 2019 Conference of the North American Chapter of the Association for Computational Linguistics: Human Language Technologies,*

Table 7: Experimental results of various toxic content detection methods across different numbers of labeled texts($K$) in multiple datasets, using RoBERTa as the basic model. The Average value represents the mean of the three baseline models. The $\Delta$ value indicates the difference between U-GIFT and the Average of the baseline models.

| Datasets | Method | Metric | K=2 | K=5 | K=10 | K=20 | K=50 | K=100 | K=500 | K=1000 |
|---|---|---|---|---|---|---|---|---|---|---|
| Jigsaw | RGFS-CA | Acc | 37.33 | 47.73 | 41.60 | 44.62 | 64.53 | 44.50 | 32.40 | 36.23 |
| | | F1 | 47.80 | 53.47 | 43.07 | 65.07 | 65.00 | 55.00 | 48.00 | 49.73 |
| | RGFS-SA | Acc | 32.40 | 33.00 | 42.00 | 60.80 | 34.20 | 60.10 | 43.40 | 35.40 |
| | | F1 | 48.00 | 49.33 | 42.60 | 63.03 | 52.00 | 60.75 | 53.20 | 52.00 |
| | Multi-Task | Acc | 39.59* | 69.43* | 79.61* | 85.15* | 89.58* | 91.06* | **93.97** | **94.39** |
| | | F1 | 52.17* | 70.25* | 79.75* | 85.17* | 89.58* | 91.08* | **94.00** | **94.42** |
| | Average | Acc | 36.44 | 50.05 | 54.40 | 63.52 | 62.77 | 65.22 | 56.59 | 55.34 |
| | | F1 | 49.32 | 57.68 | 55.14 | 71.09 | 68.86 | 68.94 | 65.07 | 65.38 |
| | U-GIFT | Acc | **52.42** | **73.58** | **85.44** | **88.20** | **91.80** | **92.85** | 93.92* | 94.25* |
| | | F1 | **45.91** | **73.53** | **85.36** | **88.14** | **91.77** | **92.81** | 93.87* | 94.21* |
| | $\Delta$ | Acc | +15.98 | +23.53 | +31.03 | +24.68 | +29.03 | +27.63 | +37.33 | +38.91 |
| | | F1 | -3.42 | +15.85 | +30.22 | +17.05 | +22.91 | +23.86 | +28.81 | +28.83 |
| HateXplain | RGFS-CA | Acc | 49.13 | 37.13 | 50.33 | 43.27 | 56.03 | 60.27 | 65.23 | 36.53 |
| | | F1 | 54.73* | 59.10* | 50.57 | 54.60 | 56.10 | 63.23 | 67.23 | 63.60 |
| | RGFS-SA | Acc | 47.50 | 52.73* | 47.70 | 52.63 | 60.97 | 62.90 | 65.00 | 72.00 |
| | | F1 | 52.83 | **59.77** | 49.00 | 55.07 | 61.23 | 63.40 | 66.50 | 72.80 |
| | Multi-Task | Acc | 55.64* | 46.07 | **59.14** | 58.60* | 70.81* | 75.50* | **82.91** | **84.40** |
| | | F1 | **56.25** | 50.75 | **59.25** | 58.75* | 71.25 | 75.67* | **82.92** | **84.42** |
| | Average | Acc | 50.76 | 45.31 | 52.39 | 51.50 | 63.53 | 66.22 | 71.05 | 64.31 |
| | | F1 | 54.60 | 56.54 | 52.94 | 56.14 | 63.64 | 67.43 | 72.22 | 73.61 |
| | U-GIFT | Acc | **56.25** | **55.50** | 58.25* | **63.08** | **70.92** | **75.75** | 76.92* | 80.50* |
| | | F1 | 53.43 | 52.63 | 57.85* | **60.96** | **70.77** | **75.71** | 76.77* | 80.46* |
| | $\Delta$ | Acc | +5.49 | +10.19 | +5.86 | +11.58 | +7.39 | +9.53 | +5.87 | +16.19 |
| | | F1 | -1.17 | -3.91 | +4.91 | +4.82 | +7.14 | +8.28 | +4.55 | +6.85 |

Table 8: Experimental results for unbalanced and unlabeled Jigsaw data (K = 10) using RoBERTa.

| | Metric | Non-toxic:Toxic 2:1 | 5:1 | 10:1 | 20:1 | 50:1 | 100:1 | 1:0 |
|---|---|---|---|---|---|---|---|---|
| Basic Model | Acc | 78.50 | 75.88 | 80.50 | 79.25 | 81.00 | 80.50 | 80.25 |
| | F1 | 78.17 | 75.22 | 80.45 | 79.11 | 80.92 | 80.44 | 80.18 |
| U-GIFT | Acc | **87.75** | **84.13** | **86.38** | **85.38** | **86.13** | **86.25** | **84.50** |
| | F1 | **87.75** | **83.99** | **86.38** | **85.36** | **86.12** | **85.73** | **84.47** |
| $\Delta$ | Acc | +9.25 | +8.25 | +5.88 | +6.13 | +5.13 | +5.75 | +4.25 |
| | F1 | +9.58 | +8.77 | +5.93 | +6.25 | +5.20 | +5.29 | +4.29 |

*Volume 1 (Long and Short Papers)*, Jill Burstein, Christy Doran, and Thamar Solorio (Eds.). Association for Computational Linguistics, Minneapolis, Minnesota, 1415–1420.

[54] Ziqi Zhang, David Robinson, and Jonathan Tepper. 2018. Detecting Hate Speech on Twitter Using a Convolution-GRU Based Deep Neural Network. In *The Semantic Web*, Aldo Gangemi, Roberto Navigli, Maria-Esther Vidal, Pascal Hitzler, Raphaël Troncy, Laura Hollink, Anna Tordai, and Mehwish Alam (Eds.). Springer International Publishing, Cham, 745–760.

Table 9: Performance assessment of our method on datasets in multiple non-English languages (K = 50, 20) using XLM-R.

| Method | Metric | SWSR (Chinese) | | Offenseval_2020 (Arabic) | | Korean-Hate-Speech (Korean) | | RP-Crowd (German) | |
|---|---|---|---|---|---|---|---|---|---|
| | | 50 | 20 | 50 | 20 | 50 | 20 | 50 | 20 |
| Basic Model | Acc | 67.25 | 63.50 | 73.25 | 63.25 | 66.50 | 62.25 | 57.50 | 55.50 |
| | F1 | 66.97 | 62.77 | 72.88 | 63.25 | 65.73 | 59.34 | 57.07 | 55.34 |
| U-GIFT | Acc | **74.00** | **66.75** | **77.25** | **69.50** | **72.00** | **71.00** | **61.00** | **58.00** |
| | F1 | **73.98** | **66.06** | **77.22** | **69.22** | **71.48** | **70.86** | **60.78** | **57.85** |
| Δ | Acc | +6.75 | +3.25 | +4.00 | +6.25 | +5.50 | +8.75 | +3.50 | +2.50 |
| | F1 | +7.01 | +3.29 | +4.34 | +5.97 | +5.75 | +11.52 | +3.71 | +2.51 |

Table 10: A series of ablation experiments on LLMs with different label sample sizes ($K$), using Llama2-7B as an example, illustrates the contributions made by the semi-supervised framework augmented by U-GIFT.

| | Metric | $K$ | | | | | |
|---|---|---|---|---|---|---|---|
| | | 5 | 10 | 20 | 50 | 100 | 1000 |
| Llama2-7B | Acc | 42.16 | 57.84 | 52.94 | 65.68 | 76.47 | 93.25 |
| | F1 | 39.08 | 55.21 | 52.41 | 65.59 | 76.36 | 93.24 |
| Llama2-7B+SSF(U-GIFT) | Acc | **50.00** | **67.65** | **72.55** | **77.45** | **89.22** | **93.83** |
| | F1 | **48.62** | **66.81** | **72.45** | **77.37** | **89.20** | **93.81** |
| Δ | Acc | +7.84 | +9.81 | +19.61 | +11.77 | +12.75 | +0.58 |
| | F1 | +9.54 | +11.61 | +20.04 | +11.78 | +12.84 | +0.57 |

[55] Botong Zhao, Yanjie Wang, Keke Su, Hong Ren, and Xiyu Han. 2023. Semi-supervised pedestrian re-identification via a teacher–student model with similarity-preserving generative adversarial networks. *Applied Intelligence* 53, 2 (2023), 1605–1618.
[56] Xianbing Zhou, Yang Yong, Xiaochao Fan, Ge Ren, Yunfeng Song, Yufeng Diao, Liang Yang, and Hongfei Lin. 2021. Hate speech detection based on sentiment knowledge sharing. In *Proceedings of the 59th Annual Meeting of the Association for Computational Linguistics and the 11th International Joint Conference on Natural Language Processing (Volume 1: Long Papers)*. 7158–7166.